\documentclass[12pt,preprint]{aastex}

\begin{document}

\slugcomment{submitted to: {\it Proceedings of the Astronomical Society of the Pacific}}

\title{Commissioning of the Dual-Beam Imaging Polarimeter for the UH 88-inch telescope}
\author{Joseph Masiero\altaffilmark{1}, Klaus Hodapp\altaffilmark{1}, Dave Harrington\altaffilmark{1}, Haosheng Lin\altaffilmark{1}}

\altaffiltext{1}{Institute for Astronomy, University of Hawaii, 2680 Woodlawn Dr, Honolulu, HI 96822, {\it masiero, hodapp, dmh, lin@ifa.hawaii.edu}}

\begin{abstract}
In this paper we present the design, calibration method, and initial results
of the Dual-Beam Imaging Polarimeter (DBIP).  This new instrument is designed
to measure the optical polarization properties of point sources, in
particular Main Belt asteroids.  This instrument interfaces between the Tek
$2048\times2048$ camera and the University of Hawaii's 88-inch telescope, and
is available for facility use.  Using DBIP we are able to measure linear
polarization with a 1-sigma Poisson signal noise of $0.03\%$ per measurement
and a systematic error of order $0.06\%\pm0.02\%$.  Additionally, we discuss
measurements of the polarization of the asteroid 16 Psyche which were taken
as part of the instrument commissioning.  We confirm Psyche's negative
polarization of $-1.037\%\pm0.006\%$ but find no significant modulation of
the signal with rotation above the $0.05\%$ polarization level.

\end{abstract}


\section{Introduction}
Optical polarimetry of asteroids enjoyed a rush of activity in the 1960s and
1970s until the limits of the optics and imagers were reached.  A resurgence
of interest in polarimetric imaging of asteroids then appeared, fueled by
previous successes with optics such as retarders and Savart plates
(e.g. \citet{serkowski74}) as well as improved CCD technology.  Polarimetric
studies of asteroids require large amounts of telescope time spread out over
many nights to fully cover rotation periods and phase angles.  The University
of Hawaii's 88-inch telescope offers our study the best access to a wide
range of observing nights, but does not have a high precision polarimeter.
We present the design and calibration results for a new instrument: the
Dual-Beam Imaging Polarimeter (DBIP) for the UH 88-inch telescope which is
capable of simultaneously measuring both orthogonal polarization states for
point sources to an accuracy of better than $0.1\%$.  The observing scheme
has been designed to make these measurements independent of flat field
effects as well as changes in seeing or extinction.  DBIP was commissioned in
March of 2007 in half-wave mode, allowing measurement of linear polarization
(Stokes $Q \& U$ vectors).  Two polarized standards and two unpolarized
standards were observed, as well as $6.8$ hours of coverage of the asteroid
16 Psyche, which has shown weak variations in polarization in the past
\citep{broglia92}.  In August 2007, DBIP will be commissioned in Full Stokes
mode, providing sensitivity to Stokes $Q, U, \& V$ while still retaining flat
field and temporal independence.

\section{Instrument Design}
\label{design}

The optical design of DBIP centers on a double-calcite Savart plate.  Calcite
splits an incident beam into two parallel beams with orthogonal
polarizations, called the ordinary (o) and extraordinary (e) beams.  The
e-beam has a slightly longer optical path in the calcite, producing
differential aberrations.  A Savart plate is two calcite blocks bonded with
the e and o beams reversed, ensuring that the two beams have the same optical
path length and same aberrations. The Savart plate used for DBIP was
originally used for Near-IR imaging polarimetry \citep{savart}, but maintains
its polarization properties in the optical regime.  It has a beam separation
of $\sim1~$mm, which scales to $6.9~$arcseconds ($31.4~$pixels) on the Tek
CCD.  As currently mounted the two beams are rotated slightly clockwise of a
North-South orientation, however we will refer to them as the northern and
southern beams throughout this paper.

Because of this relatively small separation, DBIP is most useful for
measuring point source polarization, and is incapable of measuring field
polarization (including flat field polarization).  However, as will be
discussed in \S~\ref{reduction}, the power of a dual-beam system is that when
used correctly the polarization values determined are independent of both
flat field and temporal variations.

By rotating the Savart plate, all linear polarization angles could be
explored, however this would mean that the direction of separation of the
images would be dependent on the angle being investigated, which negates the
flat-field independence provided by the Savart.  Instead, a half-wave plate
is inserted just before the Savart plate allowing the plane of polarization
of the incident beam to be rotated.  Once retarded by the half-wave plate,
the transmitted intensity of the two beams exiting the Savart plate are

\[I'_1=\frac{1}{2}\left(I+Q\cos\left({4\theta}\right)+U\sin\left({4\theta}\right)\right)\]
\[I'_2=\frac{1}{2}\left(I-Q\cos\left({4\theta}\right)-U\sin\left({4\theta}\right)\right)\]

where $I$ is the total intensity, $Q$ is the Stokes Q vector amplitude, $U$
is the Stokes U vector amplitude, $\theta$ is the angle of rotation of the
waveplate, and $I'$ is the throughput light amplitude in each of the two
split beams.

Thus by rotating the half-wave plate any angle of polarized light can be
sampled by the Savart plate while the image separation and orientation
remains constant.  There are a few different designs for half-wave plates
\citep{PanWave,GoodWave}, but a true zero-order retarder was chosen because
they tend to have fairly flat retardance across a broad range of wavelengths
and are available off-the-shelf for small (e.g. $1~$inch) diameters.  Our
half-wave plate is a birefringent polymer supplied by the Bolder Vision Optik
company and provides a retardance of $0.50\pm0.01$ wavelengths over
the$400-700~$nm wavelength range\footnote{quote from vendor's website:
http://www.boldervision.com/achro.html}.  Beginning with the expansion from
the Muller matrix for a simple retarder-polarizer setup, we have the equation:  

\[I'=\frac{1}{2}\left(I+\frac{1}{2}[(1+\cos\phi)+(1-\cos\phi)\cos(4\theta)]~Q + \frac{1}{2}\sin(4\theta)(1-\cos\phi)~U - \sin(2\theta)\sin\phi~V\right)\]

where $\phi$ is the retardance and $\theta$ is the angle of waveplate
rotation.  If the retardance varies between $0.49$ and $0.51~$wavelengths,
this means that $\phi$ varies between $176.4^\circ<\phi<183.6^\circ$.  Using
these maximal values for $\phi$ and a waveplate rotation angle of
$\theta=45^\circ$, the transmitted intensity will be:

\[I'=\frac{1}{2}\left(I-0.998~Q-0.063~V\right)\]

which corresponds to a $\le0.2\%$ depolarization of the incident polarization
(compared to the nominal $I'=0.5*(I-Q)$ for an ideal half-wave plate), as well
as some separation of Stokes $V$ into the positive and negative beams.  The
Savart plate is insensitive to Stokes $V$ and thus this term can be
neglected.  It can be shown that Stokes $U$ will have depolarization
systematics of the same order ($\sim0.4\%$).  Thus for a $10\%$ polarized
standard star we expect any retardance variation to manifest as a difference
of $\sim0.03\%$ in the measured percent polarization.  A source with
$\sim1\%$ polarization (e.g. asteroids near their maximum negative
polarization phase angle) will be measured as $\sim0.003\%$ more depolarized
than it actually is.  Both of these (as shown below) fall well within our
current limits for systematic polarization effects.  This systematic
depolarization would affect absolute polarization measures, and can be
calibrated out with a large number of polarized standard star observations.
This should not affect relative polarization measurements using a consistent
filter and waveplate rotation angles.

\begin{figure}
\begin{center}
\includegraphics[scale=0.6]{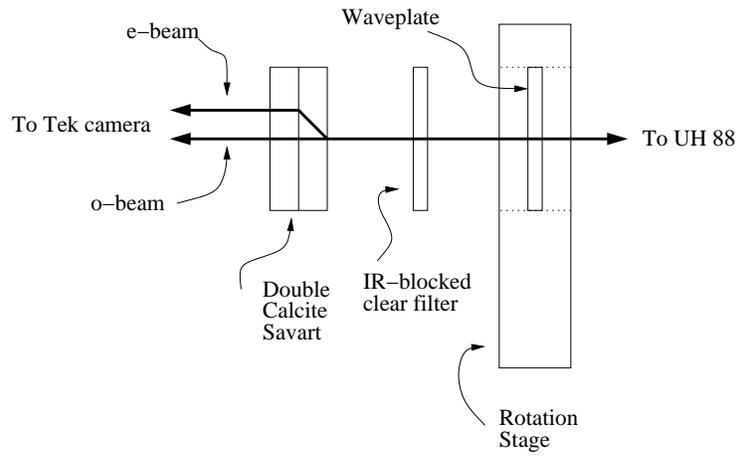}
\protect\caption{
The arrangement of optical elements in DBIP.
}
\label{fig.optics}
\end{center}
\end{figure}

An IR-blocked clear filter is chosen to restrict the observations to optical
wavelength bands.  The filter was supplied by the Custom Scientific company
and has $>90\%$ transmission in the $400-700~$nm range (typically $95\%$) and
no throughput beyond that range.  This is the closest analog to a Sloan {\it
g'+r'} filter that could be found both off-the-shelf and reasonably priced.
Ideally the filter would be located after the Savart plate along the optical
axis to prevent stray polarization signals from it, however due to physical
space constraints within the mount it could only be located between the
half-wave plate and the Savart plate.  Should the filter generate any false
linear polarization signal, unpolarized standard star calibrations should be
able to detect this and remove it from measurement results.  The biggest
problem that could result from this placement of the filter would be if the
filter had a tendency to act like a retarder.  This could cause low levels of
systematic depolarization, which would show up in the measurements of the
absolute polarization of the polarized standard stars.  A diagram of the
optical configuration can be seen in Fig~\ref{fig.optics}.

We chose to optimize DBIP for polarization analysis of point sources, and so
wide field capabilities are not required.  This meant we are not restricted
by optical elements that vignetted the field of view and thus used less
expensive, off-the-shelf components when available.  Our $1~$inch diameter
($0.8~$inch clear aperture) half-wave plate is significantly smaller than the
Tek CCD, which causes the field of view to be vignetted from the standard
$\sim7~$arcmin-sided square to a $2.5~$arcmin clear aperture.  A benefit of
this is that DBIP only requires the central $1050\times1050$ pixels to be
read out, reducing readout time to $\sim15~$seconds. 

To make this design as cost- and time-efficient as possible, we chose to use
as many existing materials as possible.  To mount our optics to the
telescope, we used a mount fabricated for the QUIRC camera \citep{quirc}.
This mount attaches directly to the UH 88-inch telescope guider and only
required minor modification to support the Tek camera.  A spacer was made to
hold the Savart plate, a bridge to hold the shutter, and the removable stage
was altered to support the rotation stage and filter.  We used a Newport PR50
series rotation stage with the accompanying SMC100 controller to rotate the
waveplate.  The PR50 model has a built-in rotary encoder allowing for
tracking of the absolute position of the rotation stage, as opposed to open
loop models which can only track the relative position to the device.  The
rotation stage has an angular resolution of $0.01^\circ$ and a guaranteed
absolute motion accuracy of $0.1^\circ$.

\begin{figure}[b]
\begin{center}
\includegraphics[scale=1.0]{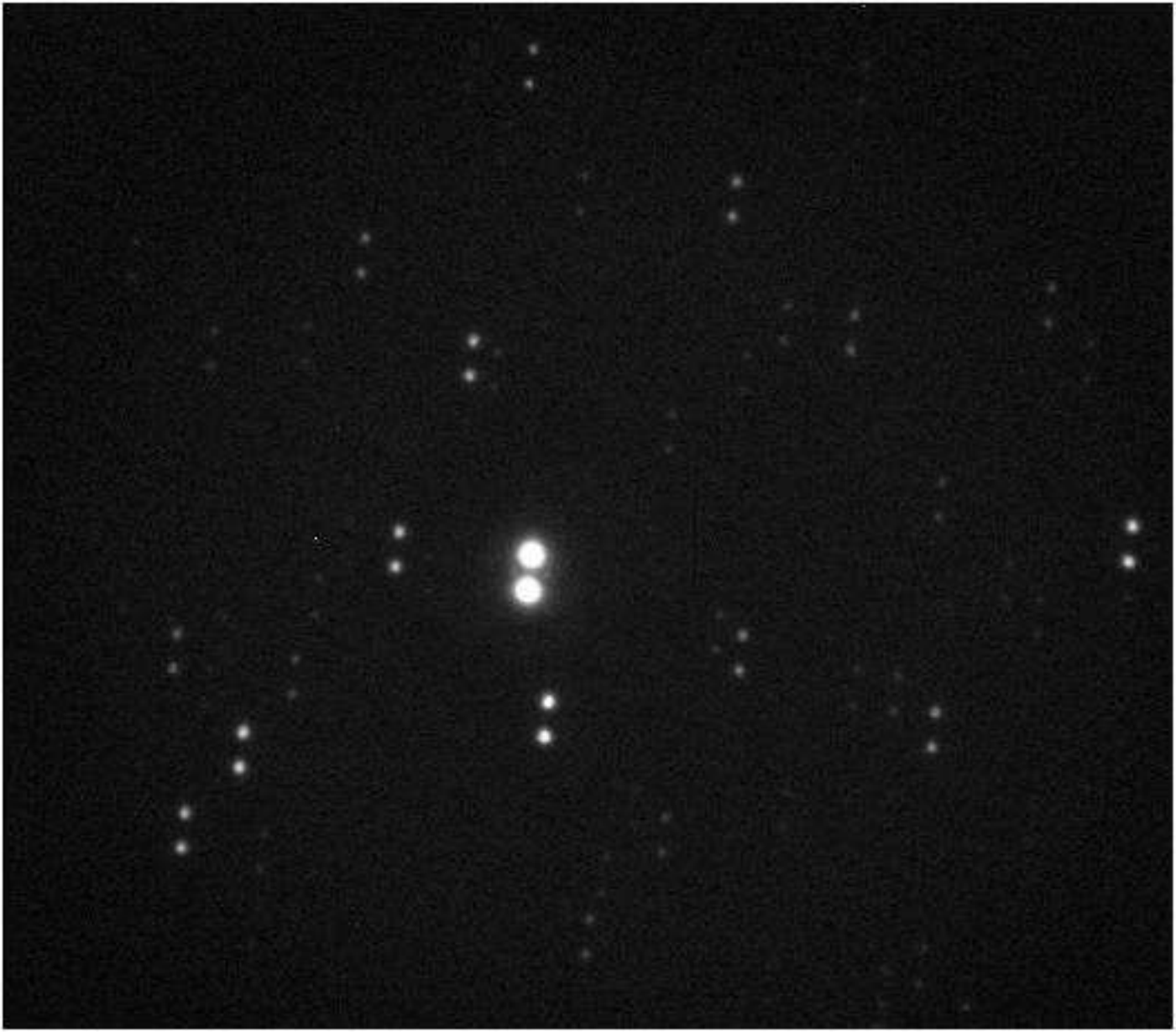}
\protect\caption{
The first-light image from DBIP.  The bright object near the center is HD 64299.  The dual images of each objects are separated by $6.9''$, and the entire field of view spans $2.5'$.  In this image, North is up and East is left.
}
\label{fig.firstlight}
\end{center}
\end{figure}

The first-light image from DBIP is shown in Fig~\ref{fig.firstlight}.  The
nature of the split image, as well as the separation and entire field of view
of the instrument can be seen in this picture.

\section{Calibration}
\label{calib}


One of the biggest challenges in calibrating DBIP was finding published lists
of polarized and unpolarized standard stars that would not saturate the
detector in less than a second, i.e. a magnitude limit of $V>9.7~$mag.  Most
lists of polarized standards are brighter than this limit, and thus most
useful for spectropolarimetry
(e.g. Keck/LRISp\footnote{http://www2.keck.hawaii.edu/inst/lris/polarimeter/polarimeter.html};
Subaru/FOCAS\footnote{http://www.naoj.org/Observing/Instruments/FOCAS/pol/calibration.html};
{\it Hubble} \citep{hubbleSTD2}).  \citet{fossati07} provide a list of polarized
standards that fall within our required magnitude range, however most of
these are in the Southern Hemisphere, and thus difficult to observe from
Mauna Kea.  From their list we are able to find two unpolarized standards
adequately observable from our location, as well as two standards polarized
at the $4\%$ and $10\%$ level.  When the peak pixel value of each image of
the target is kept near 45,000 counts the total beam will include
$\sim1.2\times10^6$ counts, which will have a 1-sigma Poisson noise of
$\sim0.09\%$.  This flux level can be reached for a V$=10.79~$mag target in
$\sim3~$seconds, or $\sim9~$seconds for a V$=12.17~$mag target.  In terms of
faint limits, DBIP can reach this flux level for a V$=17.2~$mag source with a
$15~$min exposure time.  Once both beams of the four images required to
determine the linear polarization are combined, DBIP can measure percent
polarization with a photon noise error of $\sim0.03\%$ on an individual
measurement.  Multiple sets of observations allow us to reach Poisson noise
errors of $\sim0.015\%$ for our standard stars.

Target information, as well as literature and observed values for the percent
polarization ($\%~$Pol) and the angle of polarization ($\theta$) are given in
Table~\ref{tab.polstd}.  The difference between the observed and literature
position angles are also given.  It should be noted that our measurements
show that HD 64299 is polarized at the $0.1\%$ level.  HD 64299 was first
reported as an unpolarized standard by \citet{krautter}, with a polarization
of $0.05\%\pm0.10\%$.  \citet{hubbleSTD} report HD 64299 as an unpolarized
standard with polarization (in the $B$ filter) of $0.151\%\pm0.032\%$.  These
values are consistent (within $2\sigma$) with our observed measurement of
$0.1\%\pm0.01\%$ polarization, however we are unable to determine at this
point whether this polarized signal is due to low level systematics or is a
true signal.  Note that the errors given are for Poisson noise only.
Systematic errors in our instrument are discussed below.  \citet{wardle74}
discuss in their appendix the effect of the non-Gaussian distribution of
errors on the measurement of percent polarization and angle of polarization.
These effects are important for low-to-moderate signal-to-noise ratios,
however our observations are at a S/N$\approx3500$, well above this regime.
Thus these effects have not been incorporated in our quoted errors of $\%~$P.

Based on our observations of the two polarized standard stars, we find an
offset in the angular alignment of the optics to be $9.23^\circ\pm0.32^\circ$
in the positive direction (East-of-North), where the error in this
measurement is dominated by the errors on the literature values of angle of
polarization.  This offset of $9.23^\circ$ should be subtracted from all
angular measurements before analysis.  Our observations of WD 1615-154 result
in a percent polarization measurement of $0.02\%\pm0.02\%$, consistent with a
zero result, and indicative of low-to-no systematic increase in polarization
due to the optics.

\begin{figure}[t]
\begin{center}
\includegraphics[angle=-90,width=\textwidth]{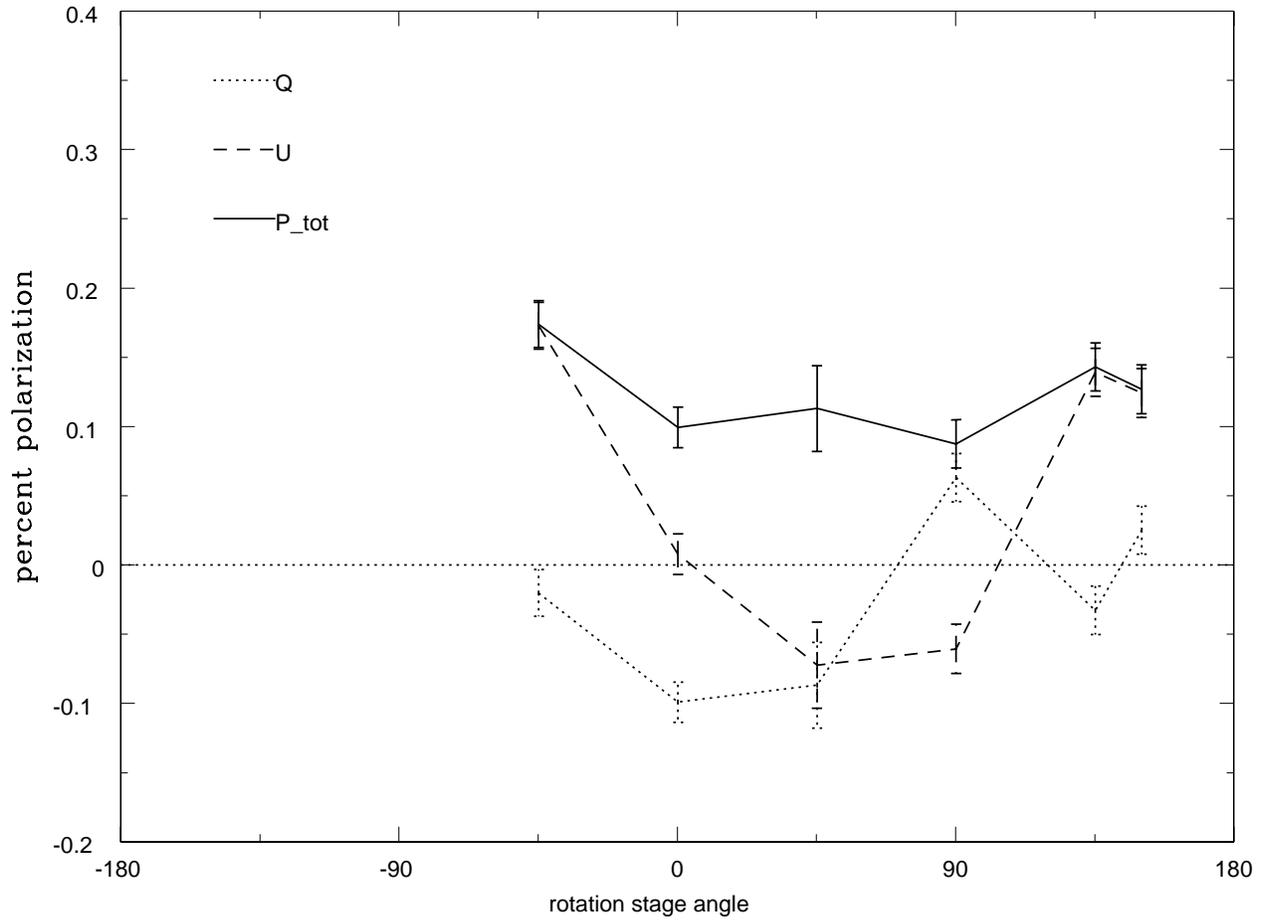}
\protect\caption{
Measurements of the polarization of HD 64299 across a range of Cassegrain rotation angles.  $Q$ and $U$ measurements are with respect to the sky.
}
\label{fig.unpol}
\end{center}
\end{figure}

\begin{figure}[b]
\begin{center}
\includegraphics[angle=-90,width=\textwidth]{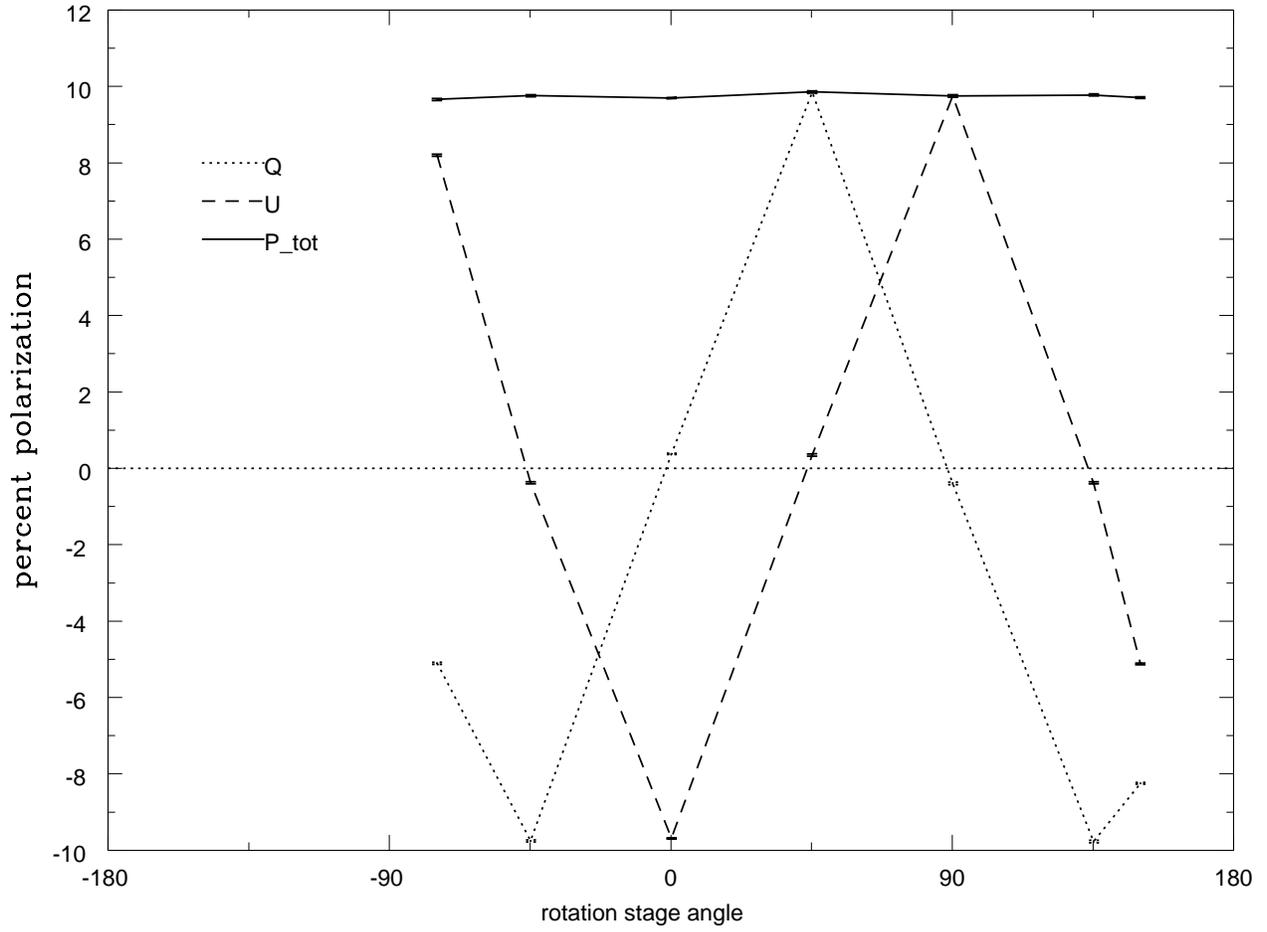}
\protect\caption{
Measurements of the polarization of NGC 2024-1 across a range of Cassegrain rotation angles.  $Q$ and $U$ measurements are with respect to the sky.
}
\label{fig.pol}
\end{center}
\end{figure}

The Cassegrain stage on UH 88-inch telescope has the ability to rotate,
allowing the angle of the instrument on the sky to be changed.  Restrictions
due to cabling only allow coverage over the range of $-75^\circ$ to
$150^\circ$ from North, but because of the $180^\circ$ symmetry in the
polarization vectors, this is sufficient to develop a full model of the
polarization with instrument rotation, and thus investigate any residual
polarization signal imparted by the telescope's primary and secondary
mirrors.  We took measurements of HD 64299 and NGC 2024-1 across the range of
instrument rotation angles to test for this signal.  Figures \ref{fig.unpol}
and \ref{fig.pol} show the changes in the fractional polarization of $Q$,
$U$, and the total polarization $P$ for various Cassegrain stage rotation
angles for HD 64299 and NGC 2024-1 respectively.  Because the polarization
signal from HD 64299 was seen to be fixed with respect to the sky when the
Cassegrain stage was rotated, and not with respect to the instrument, the
measured polarization is not due to an inherent systematic error created by
one of the optical elements (e.g. the filter).  In order to differentiate a
polarization induced by the telescope's mirrors with a true signal,
observations of a number of polarized standards with a range of position
angles is required.  This will be the goal of a future calibration campaign
using this instrument.

We find that there are variations in the polarization measurement with
rotation, with an amplitude of $0.08\%\pm0.03\%$ for a $\sim10\%$ polarized
source and $0.04\%\pm0.02\%$ for a $\sim0.1\%$ polarized source.  This
systematic error, however, does not change smoothly with instrument angle,
nor are the variations the same for both targets.  Thus until we perform
further observations we can only quote an estimated amplitude of order
$0.06\%\pm0.02\%$ for our systematic errors.  This error includes effects
from the telescope mirrors as well as from any misalignment in the optics or
dust on the optical surfaces.  This source of error is larger than the photon
Poisson error, and will affect absolute polarization measurements and limit
our accuracy when compared with literature values, however relative
polarization measurements should be unaffected by this error.  Additionally,
we see no systematic depolarization that would indicate retarder-like
behavior from the filter or deviation from $0.5~$wave retardance in the
waveplate at a level greater that the above quoted limits.

\section{Image Reduction}
\label{reduction}

\begin{figure}[t]
\begin{center}
\includegraphics[angle=90,width=\textwidth]{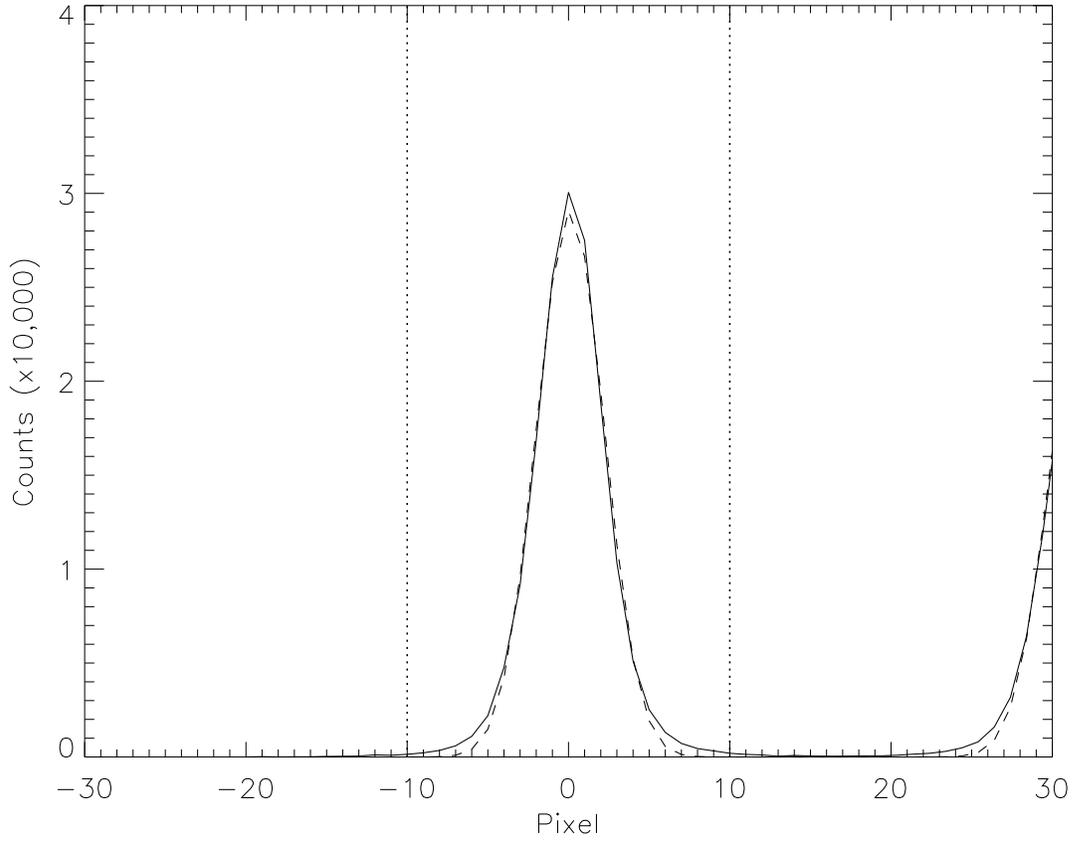}
\protect\caption{
Typical PSF for a target object centered at 0, with its orthogonally polarized component visible beyond 20.  The solid line shows example data and the dashed line indicates a best-fit 2D Gaussian model.  The dotted vertical lines indicate the region the flux is summed over for our calculations.
}
\label{fig.psf}
\end{center}
\end{figure}

\begin{figure}[b]
\begin{center}
\includegraphics[angle=90,width=\textwidth]{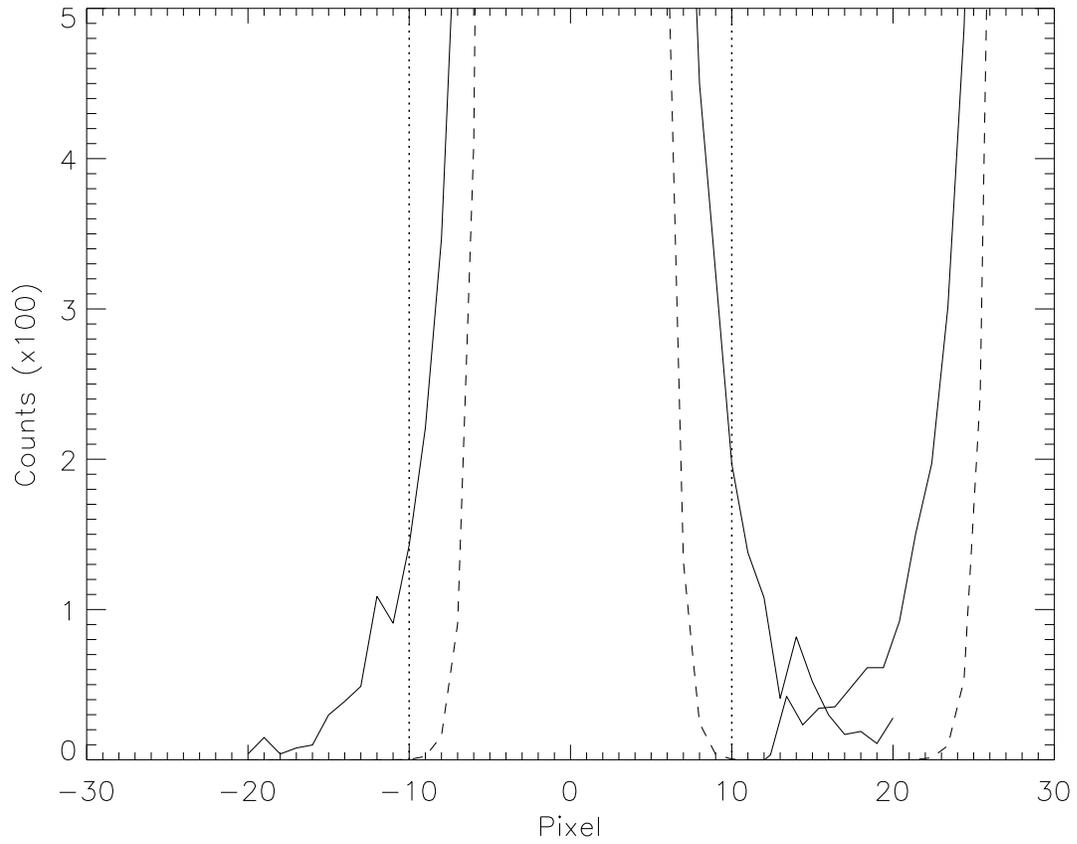}
\protect\caption{
The same as Fig~\ref{fig.psf}, but zoomed in to show detail at the edges of the wings.  Note that the $20\times20$ pixel box includes more flux than the 2D Gaussian PSF model, while avoiding any contamination from the orthogonal component.
}
\label{fig.zoom}
\end{center}
\end{figure}

Analysis of the data taken during the commissioning run is done using IDL.
Images are loaded in blocks of four (Q1, Q2, U1, U2) matching the pattern
they were obtained by the camera control script.  The designations Q1, Q2,
U1, and U2 represent, respectively, waveplate rotation angles of $0^\circ$,
$45^\circ$, $22.5^\circ$, and $67.5^\circ$ with respect to the Savart plate
orientation.  The user is then required to click on the Northern-most image
of the target in the displayed frame, this being the only user-dependent
section of the reduction.  The program then centers on both images of the
target, subtracts off the appropriate region of the bias image and sums the
signal in a $20\times20~$pixel box around the target.  This traces to a box
$4.4''$ on a side, which encompasses the $5\sigma$ radius of the PSF given an
worse-than-average seeing of $1''$.  The box around the orthogonal beam
begins at the $11\sigma$ radius, meaning the cross-contribution between the
beams is minimal.  Figure~\ref{fig.psf} shows a comparison between sample
data (solid line), a best-fit 2D Gaussian PSF (dashed line) and the limits of
the $20\times20~$pixel box.  Though the PSFs are aligned in the y-direction
on the CCD, we have shown cuts in the x-direction to better illustrate the
overlap between the beam-split PSFs.  In Fig~\ref{fig.zoom} we show a
close-up of the edges of the PSF wings, with both the fit and the box cut
displayed as above.  The data drop to a level comparable to the background
noise before reaching the orthogonal beam's integration box, meaning that
beam cross-talk is not a significant systematic error.  This is dependent on
both the focusing of the telescope and the seeing during the observations.
Our commissioning nights had a seeing of $1.1''$, which is worse than average
for Mauna Kea (average of $0.8''$), meaning bleeding between the PSFs will
not be a problem for most observing nights.

The background level is determined by taking the median of a box of
$60\times60$ pixels in the direction away from the oppositely polarized beam.
The error on this median is $\sim0.5$ counts per pixel or $\sim20~$counts over
the region the signal is summed over, while the Poissonian error of the
signal is of order $1500~$counts.  Thus the shot noise dominates our
photometry error.  This median background count is subtracted from each
pixel.

We then calculate the total fractional polarization for both the Q and U
states using the equations

\[q=\frac{Q1+Q2}{I1+I2}=\frac{1}{2}\left(\frac{S_{Q1p}-S_{Q2p}}{S_{Q1p}+S_{Q2p}}-\frac{S_{Q1n}-S_{Q2n}}{S_{Q1n}+S_{Q2n}}\right)\]
\[u=\frac{U1+U2}{I1+I2}=\frac{1}{2}\left(\frac{S_{U1p}-S_{U2p}}{S_{U1p}+S_{U2p}}-\frac{S_{U1n}-S_{U2n}}{S_{U1n}+S_{U2n}}\right)\]

where $S$ is the signal in an image and $p$ and $n$ subscripts refer to the
positive (Northern) and negative (Southern) images, respectively.  The
fractional polarizations $q$ and $u$ are equivalent to $Q/I$ and $U/I$
without assuming constant seeing or extinction between images.  It is trivial
to show that using this reduction method flat-field effects are canceled out,
assuming the location of the image has not changed significantly between
images and thus the flat fields at e.g. $Q1p$ and $Q2p$ are the same.

\section{Observations of 16 Psyche}
\label{psyche}
During the dates of DBIP's commissioning, the Main Belt asteroid 16 Psyche,
an M-type object, was near its phase of maximum negative polarization
\citep{broglia92, horizons}.  We observed Psyche to show an overall
polarization of $-1.037\%\pm0.006\%$, where a negative polarization value
refers to polarization parallel to the Sun-Object-Earth plane and a positive
polarization refers to polarization orthogonal to this plane.  The error
quoted here is noise-related only and does not account for any systematic
offsets.  The cause of negative polarization is second-order scattering of
the light by particles on the asteroid's surface \citep{karri89}.
Table~\ref{tab.psyche} lists the mean RA and Dec of Psyche on each night of
observing, as well as the apparent V magnitude, the mean phase angle
($\alpha$), and the measured percent polarization.

\begin{figure}
\begin{center}
\includegraphics[angle=-90,width=\textwidth]{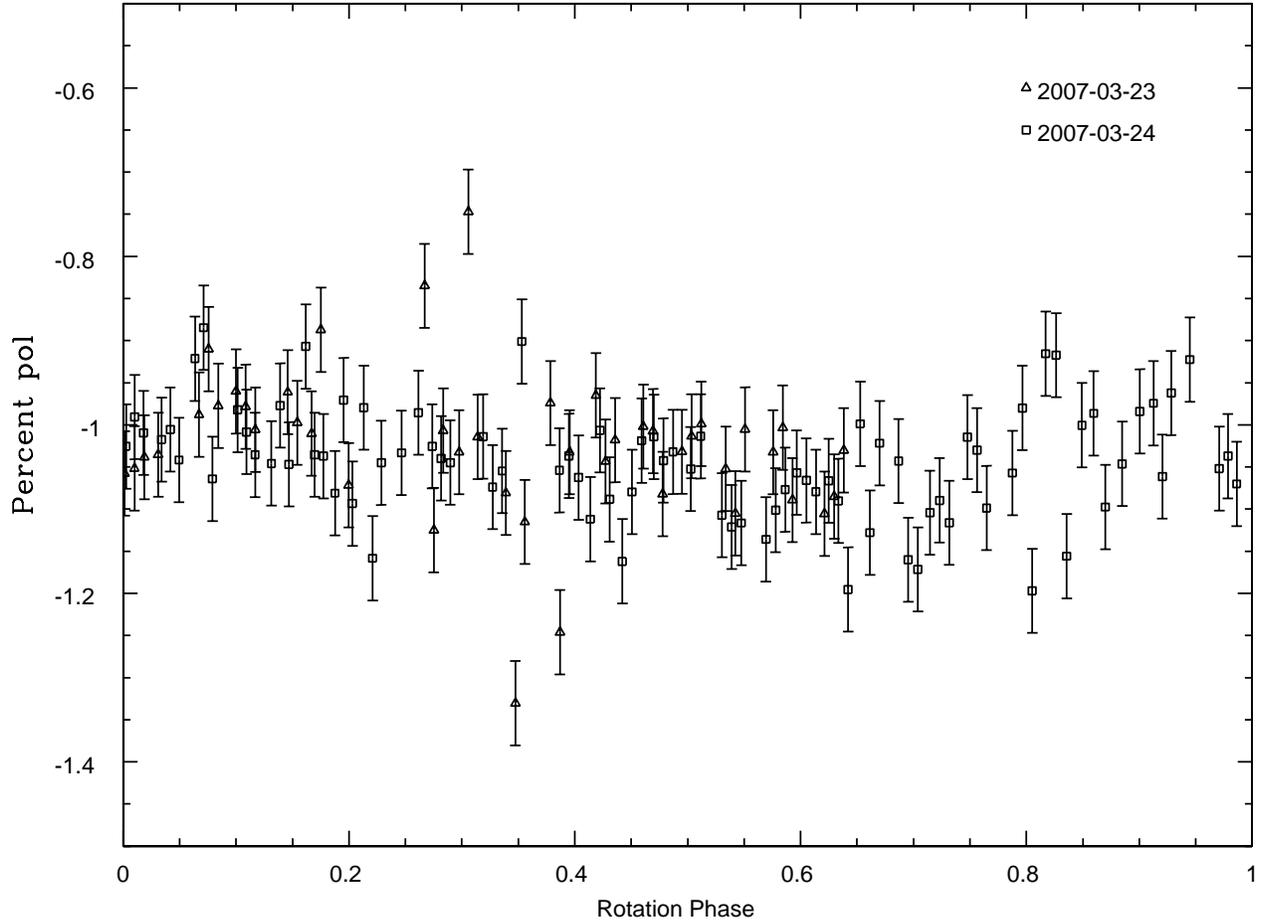}
\protect\caption{
Measurements of the polarization of 16 Psyche from the nights of 3-23-2007 and 3-24-2007.  A $0.174831~$day period is used for the rotation phase calculation.
}
\label{fig.psyche}
\end{center}
\end{figure}

Previously, \citet{broglia92} reported variations in the polarized light
curve with amplitude of $\sim0.12\%$ at a $95\%$ confidence level.  We
observed Psyche during the commissioning to verify this measurement at a
greater confidence level.  However, we are unable to detect any modulation
of Psyche's polarization with rotation phase at any confidence for a $>0.1\%$
amplitude variation.  Figure~\ref{fig.psyche} shows the data from the two
nights of observing wrapped onto a rotation period of $0.174831~$days
\citep{broglia92}.  Our observations are consistent with a constant
polarization of $-1.037\%$ and limit any amplitude of variation to $<0.05\%$.
It is possible that the discrepancy between our measurements and those of
\citet{broglia92} could be the result of changes in pole orientation.
\citet{deangelis93} measured Psyche's pole orientation to be $\lambda\sim
35^\circ$, $\beta\sim +22^\circ$ in ecliptic coordinates.  During our
observations Psyche was at an ecliptic longitude of $\lambda = 165^\circ$,
and thus a pole angle of $50^\circ$ while during the observations of
\citet{broglia92} Psyche was at an ecliptic longitude of $\lambda\sim
74^\circ$, and thus a pole angle of $109^\circ$ \citep{horizons}.  Because
our viewing orientation of Psyche is more pole-on than that of
\citet{broglia92} it is expected that we would see less variation across the
surface with rotation than those authors.  Because polarization is linked to
albedo, and thus surface properties, this may explain our null result in
looking for rotational variation in Psyche's polarization.  Conversely, if
the region of Psyche's surface that generated the polarization oscillation is
in Psyche's Southern hemisphere, it would have been observable during the
observing period of \citet{broglia92} but not during ours.

\section{Conclusions}
\label{conc}
The Dual-Beam Imaging Polarimeter, a new instrument available on the UH
88-inch telescope using the Tek camera, has been commissioned in half-wave
setup and is now available for facility use.  We find that DBIP is able to
measure percent polarization for a $V\sim10.8~$mag source with an exposure
time of $\sim3~$seconds to accuracies of $0.03\%$ from a single set of the
four observations required to determine the linear Stokes parameters.  This
level can likewise be reached for a V$\sim12.2~$mag target in $\sim9~$seconds
to a limit of V$=17.2~$mag for a 15 minute exposure time.  These measurements
are independent of flat-field, seeing, or extinction effects.  Estimates of
the systematic polarization offset due to the telescope's mirrors place the
errors at $0.06\%\pm0.02\%$.  

We set an upper limit to the rotational variation of the polarized signal of
the Main Belt asteroid 16 Psyche at $0.05\%$, much lower than previously
published results.  We postulate that this difference is due to changes in
pole orientation from previous observations to ours.

DBIP will be commissioned in full-Stokes mode during Aug 2007, when a
quarter-wave plate will be installed in addition to the half-wave plate to
provide sensitivity to both linear and circular polarization using sets of 6
images at various rotation angles for both waveplates.

\section{Acknowledgments}
The authors wish to thank Colin Aspin for providing the half-wave plate, Ed
Sousa for example control code, and Richard Shelton for help in the machine
shop.  All of the above also provided a large amount useful advice that
allowed us to bring this instrument into existence.  Additionally, we would
like to thank Rolf-Peter Kudritzki, Shadia Habbal and the entire IfA for
funding support, Nathan Huisman for computer support, and Robert Jedicke for
letting JRM take time off his thesis to get this built.  Additionally, we
thank the anonymous referee for helpful comments that greatly clarified this
paper.

\begin{deluxetable}{cccccccccc}
\tablenum{1}
\tabletypesize{\footnotesize}
\rotate
\tablecaption{Standard stars with Literature and Observed Values}
\tablewidth{0pt}
\tablehead{
\colhead{UT Obs Date}   &
\colhead{name}   &
\colhead{V mag}   &
\colhead{RA}   &
\colhead{Dec}   &
\colhead{Lit $\%~$Pol}   &
\colhead{Lit $\theta$}   &
\colhead{Obs $\%~$Pol}   &
\colhead{Obs $\theta$}   &
\colhead{$\Delta \theta$}
}
\startdata
03-23-2007 & HD 64299    & 10.11 & 07:52:25.51 & $-23^\circ 17' 46.8''$ & $0.06\pm0.13$ & n/a & $0.10\pm0.01$ & $96.95\pm4.21$ & n/a \\
03-24-2007 & WD 1615-154 & 12.40 & 16:17:55.25 & $-15^\circ 35' 52.4''$ & $0.06\pm0.24$ & n/a & $0.02\pm0.02$ & n/a & n/a \\
03-24-2007 & NGC 2024-1  & 12.17 & 05:41:37.85 & $-01^\circ 54' 36.5''$ & $9.65\pm0.06$ & $135.47\pm0.59$ & $9.70\pm0.02$ & $145.34\pm0.05$ & $9.87\pm0.59$\\
03-24-2007 & BD-12 5133  & 10.40 & 18:40:01.70 & $-12^\circ 24' 06.9''$ & $4.37\pm0.04$ & $146.84\pm0.25$ & $4.26\pm0.01$ & $155.43\pm0.10$ & $8.69\pm0.27$\\
\hline
\enddata
\vglue -0.05in
\tablecomments{
\baselineskip=0.7\baselineskip
Literature values for polarized and unpolarized standards taken from \citet{fossati07}.
}
\label{tab.polstd}
\end{deluxetable}

\begin{deluxetable}{cccccc}
\tablenum{2}
\tabletypesize{\footnotesize}
\tablecaption{Psyche Polarized Measurements}
\tablewidth{0pt}
\tablehead{
\colhead{UT Obs Date}   &
\colhead{Mean RA}   &
\colhead{Mean Dec}   &
\colhead{V mag}   &
\colhead{Mean $\alpha$}  &
\colhead{Measured $\%~$Pol}   
}
\startdata
03-23-2007 & 10:41:25 & $09^\circ 31'$ & 10.77 & $7.2^\circ$ & $-1.026\pm0.009$ \\
03-24-2007 & 10:40:45 & $09^\circ 36'$ & 10.79 & $7.5^\circ$ & $-1.048\pm0.008$ \\
\hline
\enddata
\vglue -0.05in
\label{tab.psyche}
\end{deluxetable}

\clearpage

\newpage

\newpage

\newpage

\newpage

\end{document}